\documentclass[aps,prl,twocolumn,showpacs,floatfix]{revtex4}
\usepackage{amsmath}
\usepackage{amsfonts}
\usepackage{amssymb}
\usepackage{graphicx}
\begin{document}
\title{Quantum Interferences in Cooperative Dicke Emission from Spatial Variation of Laser Phase}
\author{Sumanta Das$^1$, G. S. Agarwal$^1$, and Marlan O. Scully$^{2,3}$  }
\affiliation{$^1$Department of Physics, Oklahoma State University, Stillwater,
Oklahoma 74078, USA\\
$^2$Institute for Quantum Studies and Department of Physics, Texas A$\&$M University, Texas 77843, USA\\
$^3$Applied Physics and Materials Science Group, Eng. Quad., Princeton University, New Jersey 08544, USA}
\date{\today}
\begin{abstract}
We report generation of a new quantum interference effect in spontaneous emission from a resonantly driven system of two identical two-level atoms due to the spatial variation of the laser phase at the positions of the atoms. This interference affects significantly the spectral features of the emitted radiation and the quantum entanglement in the system. The interference leads to dynamic coupling of the populations and coherences in a basis, determined by the laser phase and represents a kind of vacuum mediated super-exchange between the symmetric and antisymmetric states.
\end{abstract}
\maketitle

Spontaneous emission from cooperative systems has been extensively
studied since the classic paper of Dicke \cite{dicke,gsab,ficek}.
The details of the emission depend on the interatomic distances
and how the system is initially prepared. The emission can further
be influenced if the system is continuously driven by a coherent
field. The two atom problem has been especially attractive in this
context as many features of cooperative emission can be analyzed
in terms of this simple problem. There is renewed interest in
these problems for quantum information sciences. Studies have
shown that spontaneous emission from cooperative systems leads to
quantum entanglement among atoms \cite{ficek}. Further with the
discovery of similarities between semiconductor quantum dots and
two level atoms \cite{barenco,gert,steel}, we have a new class of
systems where the cooperative effects can be studied in a regime
which was difficult to achieve with atoms.
In recent times such quantum dot systems are proving especially important in quantum information science \cite{vins,lukin}.\\
\indent{}In this Letter we report a new quantum interference
effect which arises from the spatial variation of the laser phase at the positions of the atoms.
We show how this phase variation affects the spectral features of the emitted radiation as well
as the quantum entanglement in the system. We further show how populations and coherences, in a basis
determined by the laser phase, get coupled dynamically. We demonstrate a kind of super-exchange
between the symmetric and anti-symmetric states and show a strong connection to the well known
vacuum induced coherence \cite{gsaprl,scully}. Further our results have implications for the decoherence of coupled qubits.\\
\begin{figure}[!h]
\begin{center}
{\includegraphics[width=6cm,height=4cm]{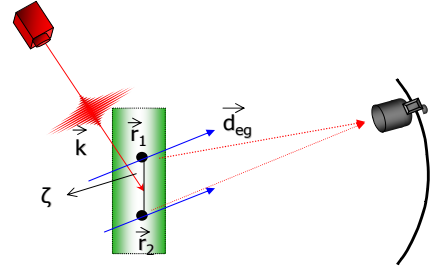}}
\caption{(Color online)Diagrammatic representation of a setup to
detect the cooperative  emission from a system of two identical
two-level atoms. The atoms are driven resonantly by a weak laser
of frequency $\omega$ and propagation vector $\vec{k}$. $\zeta$ is
the angle between the laser propagation direction and the
orientation of the atoms. $\vec{d}_{eg}$ is the dipole moment of
the atoms and $\vec{r}_{1},\vec{r}_{2}$ are the position vectors
of the atoms 1 and 2.}
\end{center}
\end{figure}
The dynamical behavior of a system of atoms undergoing cooperative
emission can be described by a master equation
approach\cite{gsab}. Let us specifically consider the system of
two identical two-level atoms with transition frequency $\omega$.
Each atom is described by the spin half angular momentum algebra.
The master equation for the dynamical behavior of this kind of a
system in the Born, Markov and rotating wave approximation is then
given by ref \cite{gsab}(pg 31-33),
\begin{eqnarray}
\frac{\partial\rho}{\partial t} & = & -i\omega\sum_{j}[S^{z}_{j},\rho] -i\sum_{j\neq k}\Omega_{jk}[S^{+}_{j}S^{-}_{k},\rho]\nonumber\\
& -& \sum_{jk}\gamma_{jk}(S^{+}_{j}S^{-}_{k}\rho-2S^{-}_{k}\rho S^{+}_{j}+\rho S^{+}_{j}S^{-}_{k}) ,\
\label{1}
\end{eqnarray}
Here (j,k = 1-2) , $\Omega_{jk}=3/2\gamma\lbrace(1-3\cos^{2}\theta)\lbrack \sin
(k_{0}r_{jk})/(k_{0}r_{jk})^2+\cos (k_{0}r_{jk})/(k_{0}r_{jk})^3\rbrack-(1-\cos^2\theta)\lbrack\cos (k_{0}r_{jk})/(k_{0}r_{jk})\rbrack\rbrace$
and $\gamma_{jk}=\gamma\lbrace
\sin(k_{0}r_{jk})/(k_{0}r_{jk})+1/2(3\cos^{2}\theta-1)\lbrack(3/(k_{0}r_{jk})^2-1)\sin(k_{0}r_{jk})/(k_{0}r_{jk})-3\cos(k_{0}r_{jk})/(k_{0}r_{jk})^2\rbrack\rbrace$
is the spontaneous decay rates for the cooperative system,  $2\gamma
= 2\gamma_{11} = 2\gamma_{22} = 4|\vec{d}_{eg}|^2\omega^{3}/3\hbar
c^{3}$ is the Einstein's A coefficient, $k_{0} = \omega/c$,
$\vec{d}_{eg}$ is the dipole moment and $\rho$ is the density operator for the
system. $\theta$ is the angle between the direction of the dipole
moment and the line joining the j\textup{th} and the k\textup{th}
atom, whose distance is denoted by $r_{jk} =
|\vec{r}_{j}-\vec{r}_{k}|$. If we assume this angle to be random,
and take an average for all possible orientations, then the
coefficients in the master equation simplify considerably and are
given by $\Omega_{jk} = -\gamma \cos(k_{0}r_{jk})/k_{0}r_{jk}$,
$\gamma_{jk} = \gamma\sin(k_{0}r_{jk})/k_{0}r_{jk}$.
The second term in the master equation (\ref{1}) is the dipole-dipole (d-d) interaction term.
It arises from the virtual photon exchange between pairs of atoms. It becomes especially
significant at small interatomic distances and has important consequences for example
it can lead to two photon resonance which was predicted and later observed experimentally \cite{gsa92}.\\
\indent{}Here we assume that the atoms are continuously driven by
a resonant laser propagating  in the direction $\vec{k}$ with
frequency $\omega$. The driving term is hence given by,
\begin{equation}
\mathcal{H}_{c} = -\hbar\sum_{j}(Ge^{i\vec{k}\cdot\vec{r}_{j}-i\omega t}S^{+}_{j}+Ge^{-i\vec{k}\cdot\vec{r}_{j}+i\omega t}S^{-}_{j}) ,
\end{equation}
where (j = 1,2) and $ G = \vec{d}_{eg}\cdot\vec{\mathcal{E}}_{o}/\hbar$ is the Rabi frequency. Note that we have included the spatially varying phase factors in the driving term. This would affect the dynamical evolution of the system. Our main focus in this Letter is to investigate new effects arising from such a phase variation. We specifically demonstrate how such phase factors can bring out new interference effects which can be experimentally investigated by studying the spectrum of the emitted radiation. While   in this letter we concentrate on spectral features and entanglement the previous papers \cite{scully} examine the effect of laser phase on emission rates. Further we specifically concentrate on the case where the relative inter-atomic distance is smaller than a wavelength when such interference are even more dramatic. The relative orientation  $\phi = \vec{k}\cdot(\vec{r}_{j}-\vec{r}_{k}) = 2\frac{\pi}{\lambda}|\vec{r}_{j}-\vec{r}_{k}|\cos\zeta$ of the two atoms and the direction of propagation of the laser drive is especially important in this context. Here $\zeta$ is the angle between the direction of the laser drive and the line joining the j\textup{th} and the k\textup{th} atom (see Fig.1). The quantum interference effects discussed in this Letter disappear if the relative orientation is perpendicular to the direction of propagation of the laser field. When the driving laser is weak, it is adequate to consider the generated states in the single photon space and clearly with two atoms starting in the ground state $|g\rangle\equiv|g_{1},g_{2}\rangle$ we would generate the symmetric state $|s\rangle$ which depends on the phase of the laser at the location of two atoms \cite{scully},
\begin{equation}
|s\rangle  \equiv \frac{1}{\sqrt{2}}( e^{i\vec{k}\cdot\vec{r}_{1}}|e,g\rangle + e^{i\vec{k}\cdot\vec{r}_{2}}|g,e\rangle).\
\end{equation}
Thus one would expect that once the system is excited to the state $|s\rangle$, it would decay to $|g\rangle$. However we show that due to quantum interferences associated with the spatial phase $\phi$, the system could also be found in the antisymmetric state $|a\rangle$ defined as ,
\begin{equation}
|a\rangle \equiv \frac{1}{\sqrt{2}}( e^{i\vec{k}\cdot\vec{r}_{1}}|e,g\rangle - e^{i\vec{k}\cdot\vec{r}_{2}}|g,e\rangle) .\
\end{equation}
Clearly, if we are working with single photon excitation then it should be adequate to deal with the states $|s\rangle , |a\rangle$ and $|g\rangle$. In order to see this we find from the master equation that the population in the symmetric state $|s\rangle$ is governed by,
\begin{eqnarray}
\dot{\rho}_{ss} & = & -2(\gamma+\gamma_{12}\cos\phi)\rho_{ss} -i\sin\phi(\gamma_{12}+i\Omega_{12})\rho_{as}\nonumber\\
& &+i\sin\phi(\gamma_{12}-i\Omega_{12})\rho_{sa}  ,\
\label{50}
\end{eqnarray}
We immediately see that the population in the symmetric state
decays at the rate $2(\gamma+\gamma_{12}\cos\phi)$,  however it is
also effected by the presence of atomic coherence terms
$\rho_{as}$ and $\rho_{sa}$ which are dynamically generated. This
coupling of populations to the coherences is at the heart of the
quantum interference phenomenon \cite{harris} that we discuss in
this letter. From Eq.(\ref{50}) it is clear that this coupling
vanishes when the laser propagates in a direction perpendicular to
the location of the two atoms$(\phi = 0)$. Further from the
structure of Eq.(\ref{50}) we can say that such quantum
interferences should be especially important for smaller
inter-atomic distances as then $\Omega_{12}$ is large and the
coherence terms strongly influence the population dynamics of the
symmetric state.\ Note further that for small times the effect of
the quantum interferences does not show up as the solution of
$\rho_{ss}$ is then,
\begin{equation}
\rho_{ss}\cong 1-2t(\gamma+\gamma_{12}\cos\phi) ,\
\end{equation}
and hence the effect of interferences should appear in physical parameters which are determined by the long time dynamics.\
\begin{figure}
\begin{center}
{\includegraphics[width=5cm, height=4cm]{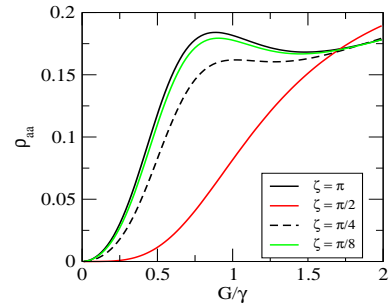}}
 \caption{(Color online) Population of the anti-symmetric state as function of the Rabi frequency for an inter-atomic distance of $\lambda/8$ and different orientation of the laser. All plotted parameters are dimensionless.}
\end{center}
\end{figure}
\begin{figure}[!h]
\begin{center}
{\includegraphics[width=5cm, height=4.5cm]{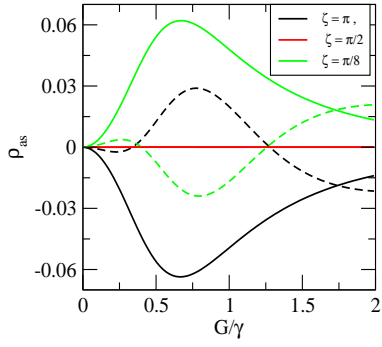}}
 \caption{(Color online) Atomic coherence $\rho_{as}$ as a function of the Rabi frequency for an inter-atomic distance of $\lambda/8$ and for different orientations of the laser. The solid and dashed lines correspond to the real and imaginary parts of $\rho_{as}$.}
\end{center}
\end{figure}
From the master equation we find that if the system starts in the
initial state $|s\rangle$, then the population $\rho_{aa}$ of the
antisymmetric state $|a\rangle$ grows as,
\begin{equation}
\rho_{aa} \sim \sin^2\phi (|\gamma_{12}+i\Omega_{12}|^2 t^2) ,
\end{equation}
Thus the states $|s\rangle$ and $|a\rangle$ get coupled by the vacuum of the electromagnetic field provided that $\phi \neq 0$ (modulo $\pi$). This is a process in which the transition $|s\rangle\rightarrow|a\rangle$  is mediated via the state $|g\rangle$. It is to be noticed that the asymmetric state for small values of the driving field remains unpopulated if $\zeta = \pi/2$ ($\phi = 0$)[see fig. 2]. However at larger values of the Rabi frequency the two photon state $|e,e\rangle$ gets populated and this changes the dynamical evolution leading to the population of the state $|a\rangle$. In Fig.(3) we show the coherence $\rho_{as}$ which is quite significant for non-zero values of the angle $\zeta$.\\
\indent{}To investigate the effects of this interference we study the steady state spectrum of cooperative emission. The incoherent part of the steady state emission spectrum integrated over all solid angles is given by,
\begin{eqnarray}
S(\omega) & = & \text {Re} \sum_{ij}\gamma_{ij}\int^{\infty}_{0}d\tau e^{-z\tau}\lim_{t\rightarrow\infty}\lbrack\langle\hat{S}^{+}_{i}(t+\tau)\hat{S}^{-}_{j}(t)\rangle \nonumber\\
& &- \langle\hat{S}^{+}_{i}(t+\tau)\rangle\langle\hat{S}^{-}_{j}(t)\rangle\rbrack_{z = i(\omega-\omega_{0})/\gamma}.\
\label{120}
\end{eqnarray}
We have calculated Eq.(\ref{120}) when the system is driven weakly by a coherent field and for small interatomic distances. Under these conditions the quantum interference effects are dominant.
The results of our numerical calculations are shown in Figs (4-5).
In the figure (4) we show the incoherent part of the normalized steady state spectrum for a weak coherent drive ($G = 0.1\gamma$) and small inter-atomic separation, $r_{12} = \lambda/8$. We have normalized the incoherent part of the steady state spectrum by dividing it with two times the steady state value of $[\langle S^{+}S^{-}\rangle - \langle S^{+}\rangle\langle S^{-}\rangle]$ for a single two level atom \cite{mol}. The spectrum exhibits a doublet structure because of the strong dipole-dipole interaction $\Omega_{12}$ for small inter-atomic distances. The quantum interferences arising from the spatial phase factor $\phi$ determine the characteristics of the doublets. For example the peak of the doublet is almost seven times greater, when $\vec{k}$ is parallel or anti-parallel to $\vec{r}_{12}$ in comparison to when $\vec{k}\perp\vec{r}_{12}$. The Fig.(5) shows the incoherent steady state spectrum for a moderately strong driving field strength ($G = 1.0\gamma$). The inset in Fig.(5) is for still larger field strength.
\begin{figure}
\begin{center}
{\includegraphics[width=5cm,height=4cm]{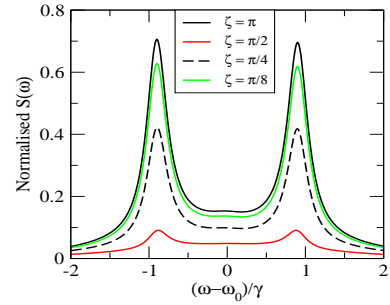}}
 \caption{(Color online)Normalized steady state spectrum of incoherent emission from
two identical two level atoms for interatomic separation of
$\lambda/8$ and Rabi frequency of 
$0.1\gamma$. The relative orientation is given by $\phi =
2\frac{\pi}{\lambda}|\vec{r}_{i}-\vec{r}_{j}|\cos\zeta$.}
\end{center}
\end{figure}
The doublet structure vanishes for moderately strong drive as seen in Fig.(5)
and we get only the broadened central peak at $\omega = \omega_{0}$.
The quantum interference leads to pronounced asymmetry in the spectrum.
For even higher field strength (inset of Fig. 5) the cooperative effects
are almost insignificant and  we get the Mollow spectrum \cite{mol} for a two level atom.\\
\indent{}The coupling of coherences to populations in the Dicke
problem of cooperative emission can be understood as vacuum
induced coherence effect\cite{gsab,ficek,scully}. This can be
appreciated more clearly at the level of Schr$\ddot{o}$dinger
equation.
\begin{figure}
\begin{center}
{\includegraphics[width=4.5cm, height=3.8cm]{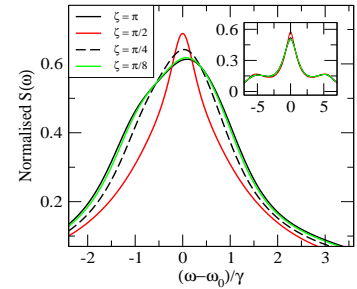}}
 \caption{(Color online) Normalized steady state spectrum of incoherent emission
 for interatomic separation of $\lambda/8$ and
 Rabi frequency G = $1.0\gamma$.
 In the inset we show the spectrum for G = 3.0$\gamma$.}
\end{center}
\end{figure}
The basic Hamiltonian between the vacuum of the electromagnetic field and the atoms  in the interaction picture can be written as,
\begin{equation}
\mathcal{H}_{I}(t) = \sum_{jks}\lbrace g_{jks}a_{ks}e^{-i\omega_{ks}t}(S^{+}_{j}e^{-i\omega t} + S^{-}_{j}e^{i\omega t}) + H.C.\rbrace ,
\end{equation}
Here $g_{jks} = -i(2\pi ck/\hbar L^3)^{1/2}(\vec{d}\cdot\hat{\epsilon}_{ks}) e^{i\vec{k}\cdot\vec{r}_{j}}$ is the vacuum coupling strength and the field annihilation(creation) operator is given by $a_{ks}$($a^{\dagger}_{ks}$). The subscripts $(k, s)$ denote the k$^{th}$ mode of the field with polarization along $\hat{\epsilon}_{ks}$. The initial state is $|s,\{0_{ks}\}\rangle$, and the final state is $|a,\{0_{ks}\}\rangle$. Iterating the Schr$\ddot{o}$dinger equation to second order in $\mathcal{H}_{I}(t)$ we find that the lowest order non-vanishing contribution to the transition amplitude is,
\begin{equation}
\frac{d}{dt}\langle a|s(t)\rangle\equiv -\frac{1}{\hbar^{2}}\lim_{t\rightarrow\infty}\int^{t}_{0}d\tau\langle a,\{0_{ks}\}|\mathcal{H}_{I}(t)\mathcal{H}_{I}(\tau)|s,\{0_{ks}\}\rangle.
\end{equation}
A long calculation then leads to,
\begin{equation}
\frac{d}{dt}\langle a|s(t)\rangle \sim i\sin\phi(\gamma_{12}+i\Omega_{12}) ,
\end{equation}
One can clearly see that this transition amplitude is zero if $\phi = 0$(modulo $\pi$). The second order transition amplitude (11) from the state $|s\rangle$ to $|a\rangle$ is mediated via the ground state $|g\rangle$. We have thus shown an intriguing connection between the quantum interference effects arising from spatial variations of the laser phase and the vacuum induced coherence effects.\\
\begin{figure}
\begin{center}
{\includegraphics[width=5cm,height=4cm]{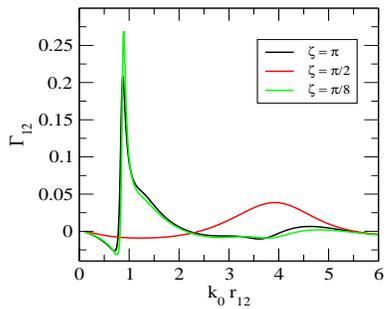}}
 \caption{(Color online) The quantum correlation between two atoms
 $\Gamma_{12} = \text{Re}[\langle \hat{S}^{+}_{1}\hat{S}^{-}_{2}\rangle/\langle \hat{S}^{+}_{1}\rangle\langle\hat{S}^{-}_{2}\rangle]-1$
 as a function of distance between the atoms for a Rabi frequency of $G =
 0.1\gamma$}
\end{center}
\end{figure}
\indent{}We conclude the Letter with a discussion of how the quantum entanglement between two atoms (qubits) also depends in a significant way on the spatial variation of the phase $\phi$.
Note that the entanglement in system arises from the fact that the density operator of the two atoms does not factorize,
$\rho \ne \rho^{(i)}\otimes\rho^{(j)}$.
This happens due to cooperative emission \cite{ficek}. The non-factorizability of the density matrix
is especially significant due to the $\Omega_{12}$ term in the dynamics. We show in the Fig.(6),
existence of the quantum correlation $\Gamma_{12} = \text {Re}[\langle\hat{S}^{+}_{1}\hat{S}^{-}_{2}\rangle/\langle\hat{S}^{+}_{1}\rangle\langle\hat{S}^{-}_{2}\rangle]-1$
for small inter-atomic distance and for different values of the angle between laser propagation direction and the line joining the two atoms. In absence of any entanglement in the system such correlation would vanish. One can see clearly from the Fig.(6) that at small interatomic separation the presence of the laser phase significantly effects the quantum correlation. Around $r_{12}\sim\lambda/6$ the value of the quantum correlation is about 25 times more in presence of the laser phase $(\zeta = \pi, \pi/8)$ in comparison to when $\phi = 0 (\zeta = \pi/2)$. Thus the quantum interference can lead to strong entanglement in the system at small interatomic separation. One can further characterize quantitatively such entanglement by calculating its concurrence.\\
\indent{} Hence we have shown how the variation of the laser phase at the positions of the atoms can lead to new quantum interference effects. The phase variation is found to affect the spectral features of cooperative emission significantly and generate strong entanglement in the system. We further demonstrate that the coupling between the symmetric and antisymmetric states has strong connection to the vacuum induced coherence in the system. A plausible system for the observation of the interference effects of this Letter would be semiconductor quantum dots. Note that the splitting in photoluminescence spectra from a system of coupled quantum dots was observed \cite{gert}.
The dots in these experiments satisfy the condition, wavelength $\gg$ interdot distance $\gtrsim$ size of the dot. Thus our theoretical results sould be observable in such systems. Further averaging over the finite size of the spatial wavefunction of the dot is expected to change the results, say for spectra, by few percent for dots of few nm size. We hope to present details in a comprehensive publication.\\
\indent{}SD and GSA  gratefully acknowlege support from NSF-Phys0653494 and NSF-CCF-0829860 and MOS thanks ONR and the Welch Foundation grant no A1261.


\begin{thebibliography}{999}
\bibitem{dicke}
R. H. Dicke , Phys. Rev. {\bf 93}, 99 (1954); J. H. Eberly and N.
E. Rehler , Phys. Rev. A {\bf 2}, 1607 (1970); R. H. Lehmberg ,
Phys. Rev. A {\bf 2}, 889 (1970); M. Macovei et. al. Phys. Rev.
Lett. {\bf 91}, 233601 (2003); ibid {\bf 98}, 043602 (2003).

\bibitem{gsab}G. S. Agarwal , Springer Tracts in Modern Physics: Quantum
Optics (Springer-Verlag, Berlin, 1974).

\bibitem{ficek}R. Tanas and Z. Ficek , J. Opt. B {\bf 6}, S90 (2004); Z. Ficek
and S. Swain , J. Mod. Opt {\bf 49}, 3 (2002); J. von. Zanthier
et. al. Phys. Rev. A {\bf 74}, 061802(R) (2006).

\bibitem{barenco}
A. Barenco et. al. Phys. Rev. Lett. {\bf 74}, 4083 (1995).

\bibitem{gert} Gert
Schedelbeck, \textit{et. al. Science} {\bf 278}, 1792 (1997); M. Bayer \textit{et. al. Science} {\bf 291}, 451 (2001); H. J. Krenner, \textit{et. al.} Phys. Rev. Lett. {\bf 94}, 057402 (2005).

\bibitem{steel} Xiaodong Xu, \textit{et. al.} \textit{Science} {\bf 317}, 929 (2007).

\bibitem{vins}
David P. DiVincenzo , \textit{Science} {\bf 270}, 255 (1995);
Daniel Loss and David P. DiVincenzo , Phys. Rev. A {\bf 57}, 120 (1998);
Xiaoqin Li, \textit{et. al. Science} {\bf 301}, 809 (2003).

\bibitem{lukin} J. R. Petta, \textit{et. al.} \textit{Science} {\bf 309}, 2180
(2005)

\bibitem{gsaprl}
G. S. Agarwal, Phys. Rev. Lett. {\bf 84}, 5500 (2000); M. Kiffner
et. al. Phys. Rev. Lett. {\bf 96}, 100403 (2006); Yaping Yang et.
al. Phys. Rev. Lett. {\bf 100}, 043601 (2008).

\bibitem{scully}
Marlan O. Scully \textit{et. al} Phys. Rev. Lett. {\bf 96}, 010501 (2006);
M. O. Scully, Laser Physics {\bf 17} 635 (2007);
A. Svidzinsky, \textit{et. al} Phys. Rev. Lett. {\bf 100}, 160504 (2008); 
Z. Ficek and B. C.  Sanders, Phys. Rev. A {\bf 41}, 359 (1990); T. G. Rudolp, Z. Ficek and B. J. Dalton, Phys. Rev. A {\bf 52}, 636 (1995).
 
\bibitem{gsa92}
G. V. Varada and G. S. Agarwal , Phys. Rev. A {\bf 45}, 6721
(1992); J. R. R. Leite and C. B. de Araujo, Chem. Phys. Lett. {\bf
73}, 71 (1980); C. Hettich \textit{et. al.} \textit{Science} {\bf
298}, 385 (2002); M. Orrit, \textit{Science} {\bf 298}, 369 (2002)

\bibitem{harris}
S. E. Harris , Phys. Rev. Lett. {\bf 62}, 1033 (1989).

\bibitem{mol}
B. R. Mollow , Phys. Rev. {\bf 188} 1969 (1969).

\end{thebibliography}
\end{document}